\documentclass[twocolumn,epjc3]{svjour3}  
\smartqed
\usepackage[utf8,latin1]{inputenc}
\newcommand{\qm}[1]{``#1''}
\usepackage{amsmath}
\usepackage{amssymb}
\usepackage{mathabx}
\usepackage{mathrsfs}  
\usepackage{cases}
\usepackage{revsymb}
\usepackage{tabularx}
\usepackage{colortbl}
\usepackage{graphicx}
\usepackage{dcolumn}
\usepackage{xcolor}
\usepackage{bm}
\usepackage{hyperref}
\hypersetup{colorlinks, linkcolor={red},citecolor={blue},urlcolor={blue}}  
\usepackage[square,numbers,sort&compress]{natbib}
\usepackage{hyperref}

  \def\nn{\nonumber} 
\newcommand{\dd}{{\rm d}}
\newcommand{\OO}{{\rm O}}

\RequirePackage{graphicx}
\journalname{Eur. Phys. J. C}

\begin{document}

\title{First post-Newtonian $N$-body problem in Einstein-Cartan theory with the Weyssenhoff fluid: equations of motion}
\titlerunning{First post-Newtonian $N$-body problem in  Einstein-Cartan theory with the Weyssenhoff fluid: equations of motion}

\author{Emmanuele Battista\thanksref{e1,e2,addr1}
\and
Vittorio De Falco\thanksref{e3,addr2,addr3}}

\thankstext{e1}{e-mail: emmanuele.battista@univie.ac.at}
\thankstext{e2}{e-mail: emmanuelebattista@gmail.com}
\thankstext{e3}{e-mail: vittorio.defalco-ssm@unina.it}

\authorrunning{Battista \& De Falco (2022)}

\institute{Department of Physics, University of Vienna, Boltzmanngasse 5, A-1090 Vienna, Austria \label{addr1}
\and
Scuola Superiore Meridionale, Largo San Marcellino 10, 80138 Napoli, Italy\label{addr2}
\and
Istituto Nazionale di Fisica Nucleare, Sezione di Napoli, Complesso Universitario di Monte S. Angelo, Via Cintia Edificio 6, 80126 Napoli, Italy \label{addr3}}

\date{Received: \today / Accepted: }

\maketitle

\begin{abstract}
We derive the equations of motion for an $N$-body system in the Einstein-Cartan gravity theory at the first post-Newtonian order by exploiting the Weyssenhoff fluid as the spin model. Our approach consists in performing the point-particle limit of the continuous description of the gravitational source. The final equations provide a hint for the validity of the effacing principle at 1PN level in Einstein-Cartan model. The analogies with the general relativistic dynamics involving the macroscopic   angular momentum are also discussed.
\end{abstract}

\section{Introduction}
\label{sec:intro}
The \emph{$N$-body problem} consists in describing the evolution of $N$ massive objects under their mutual gravitational attractive forces. If we regard the gravitational interaction \emph{\'a la Newton}, we need to solve two issues: (1) determining the equations of motion of the interacting extended bodies (represented by partial-integro differential equations); (2) solving this problem to infer their trajectories. This complex pattern can be drastically simplified if the $N$ bodies keep \emph{mutually    well separated} (i.e., their separations are greater than their typical sizes). This configuration permits to neglect, to a good approximation, the contributions ensuing from the quadrupole and higher-order multipole moments of the bodies to their external gravitational fields. Therefore, the extended objects can be modelled as $N$ point-like masses via the \emph{point-particle procedure} \cite{Poisson-Will2014}. This implies that now ordinary differential equations rule the dynamics, and numerical approaches are of fundamental importance to extrapolate the whole motion \cite{Aarseth2009}. For particular configurations it is possible to determine semi-analytical or even analytical solutions \cite{Meyer1981,Wang1991,Pupyshev1994,Goldstein2002}. 

The situation completely changes when gravity is framed in general relativity (GR), because the following complications arise: (1) \emph{non-linear geometric structure of GR}, which can spoil the well-posed mathematical formulation of the problem \cite{Ehlers1980,Bruhat1969,Bruhat2014}; (2) \emph{self-referential controversy} manifesting in the fact that the equations of motion are contained in the gravitational field equations \cite{Maggiore:GWs_Vol1,Blanchet2014}; (3) \emph{finite propagation of gravity interaction} (contrarily to the action at a distance in Newtonian physics), which yields retarded-partial-integro differential equations \cite{Maggiore:GWs_Vol1,Blanchet2014,Poisson-Will2014}.   

These conceptual and mathematical difficulties can be overcome if we exploit \emph{approximation schemes} and \emph{break the general covariance of the GR theory} by working in special classes of coordinate systems, e.g., harmonic coordinates. Simplifications occur if we assume that the \emph{gravitational source is post-Newtonian (PN)}, namely it is slowly moving, weakly self-gravitating, and weakly stressed \cite{Maggiore:GWs_Vol1,Blanchet2014}. This hypothesis permits to apply in the near zone (which covers the whole gravitational source) the \emph{PN approximation method}, where we expand the model parameters in terms of $1/c$ \cite{Maggiore:GWs_Vol1,Blanchet2014}, engendering the appearance of \emph{static potentials} without retardation effects. Finally, if the bodies are \emph{mutually well separated}, we can apply the \emph{point-particle limit}, pursuing the same strategy of classic physics. In this skeleton process, the integrals underlying basic quantities exhibit divergences exactly at the location of the particles. However, \emph{self-field regularization methods} (represented by Hadamard and dimensional techniques) are employed to heal the infinities (see Ref. \cite{Blanchet2014} and references therein for details). 

The PN approximation procedure implies that after having chosen a coordinate system, \emph{the test particles' motion occurs in the Newtonian absolute  Euclidean space}  \cite{Blanchet2014}, descending thus into the classical physical framework. Nevertheless, \emph{the equations of motion still preserve their relativistic nature}, since they remain invariant under a global PN-expanded Lorentz transformation, admit a correct perturbative limit when $N-1$ masses tend to zero, and are conservative when gravitational radiation-reaction effects are nullified \cite{Blanchet2014}.   

The PN approximation scheme was pioneered in 1917 by Lorentz and Droste, who worked out the first post-Newtonian (1PN) corrections to the Newtonian dynamics within GR \cite{Droste1917,Lorentz1937}. In 1938, Einstein, Infeld, and Hoffmann (EIH) \cite{Einstein1938,Infeld1960} re-derived these results for $N$ bodies by making use of the \emph{surface integral method}. Only in 1985, Damour and Deruelle provided the first analytical solution, expressed in a quasi-Newtonian form, to the two-body problem at the 1PN level \cite{Damour1985}. Since these first solid achievements, the theoretical progresses on the GR dynamics  attained very high PN orders via various methods. The works can be classified for non-spinning \cite{Blanchet2009a,Marchand2017,Bini2020r,Bern2021dq} and spinning \cite{Bohe2015a,Bohe2015a,Levi_2016,Cho2021} compact binary systems.  

All these developments find crucial applications in: the motion of $N$ point-like bodies for the description of planets' dynamics in the Solar System, including also the related GR effects \cite{Einstein1938,Will1993}; the gravitational radiation-reaction force in binary pulsars \cite{Weisberg2005,Weisberg2016}; the emission of gravitational waves from inspiralling compact binaries up to very high PN orders \cite{Schmidt2020,Cho2021,Cho2022}. 

In this article, we are motivated to study the $N$-body problem  in the Einstein-Cartan (EC) theory, an extension of GR where besides the curvature, which is related to the mass-energy distribution, there is also the torsion tensor, which is linked with the microscopic spin density \cite{Hehl1976_fundations}. Hereafter, the term \qm{spin} will refer to the quantum intrinsic angular momentum of bodies. This work is part of a research program aiming at modelling the gravitational-wave theory in EC geometry \cite{Paper1,Paper2}, which permits to analyze the spin contributions to gravitational phenomena. Besides the latter topic, it would be also interesting to analyse and explore some further applications of our developments in other physical contexts. Our approach relies on the same assumptions as in GR (i.e., PN source and mutually well separated bodies), but it employs the Weyssenhoff fluid \cite{Weyssenhoff1947,Boehmer2006} to treat the spin effects inside the matter. 
The article is essentially divided into three parts: derivation of the $N$-body equations of motion in EC theory at the 1PN order (see Sec. \ref{sec:EC}); applications of our findings to binary systems (see Sec. \ref{sec:binary_system}); discussion about our results and future perspectives (see Sec. \ref{sec:end}).

\emph{Notations.} We use metric signature  $(-,+,+,+)$. Greek indices take values  $0,1,2,3$, while lowercase Latin ones $1,2,3$. The determinant of the metric $g_{\mu \nu}$ is denoted by $g$. $\varepsilon_{kli}$ is the total antisymmetric Levi-Civita symbol.  The spacetime coordinates are $x^\mu = (ct,\boldsymbol{x})$. Four-vectors are written as $a^\mu = (a^0,\boldsymbol{a})$, and $\boldsymbol{a} \cdot \boldsymbol{b}:= \delta_{lk}a^l b^k$, $\vert \boldsymbol{a} \vert\equiv a :=  \left(\boldsymbol{a} \cdot \boldsymbol{a}\right)^{1/2}$, and $\left(\boldsymbol{a} \times \boldsymbol{b}\right)^i := \varepsilon_{ilk} a^l b^k$. The symmetric-trace-free projection of a tensor $A^{ij\dots k}$ is indicated with the symbol $A^{\langle ij\dots k \rangle }$. Round (respectively, square) brackets around a pair of indices stands for the usual symmetrization (respectively, antisymmetrization) procedure, i.e., $A_{(ij)}=\frac{1}{2}(A_{ij}+A_{ji})$ (respectively, $A_{[ij]}=\frac{1}{2}(A_{ij}-A_{ji})$).

\section{Post-Newtonian $N$-body problem} 
\label{sec:EC}

In this section, we first delineate briefly the Weyssenhoff fluid  in Sec. \ref{sec:Weyssenhoff_fluid}, and then  we deal with the $N$-body problem at 1PN level and the related point-particle procedure  in Sec. \ref{sec:PPL_Continuous}. 

\subsection{The Weyssenhoff fluid}
\label{sec:Weyssenhoff_fluid}

In this section, we introduce the Weyssenhoff model within the EC theory (see Sec. \ref{sec:model}) and its post-Newtonian description (see Sec. \ref{sec:PN_description}).

\subsubsection{Model and dynamics}
\label{sec:model}

The EC model is a theory of gravity defined on a four-dimensional Riemann-Cartan spacetime manifold endowed with a symmetric metric tensor $g_{\alpha \beta}$ and the most general metric-compatible affine connection $\Gamma^{\lambda}_{\mu \nu}$, whose symmetric and antisymmetric parts read as, respectively,   
\begin{subequations} 
\begin{align}
\Gamma^\lambda_{(\mu \nu)} &=   \hat{\Gamma}^{\lambda}_{\mu \nu} +2 S^{\lambda}_{\phantom{\lambda} (\mu \nu)},
\label{eq:symmetric_part}
\\
\Gamma^\lambda_{[\mu \nu]} &  := S_{\mu \nu}^{\phantom{\mu \nu} \lambda}, 
\label{eq:torsion_tensor}
\end{align}
\end{subequations}
where $\hat{\Gamma}^{\lambda}_{\mu \nu}$ corresponds to the \emph{Levi-Civita connection} and 
$S_{\mu \nu}^{\phantom{\mu \nu} \lambda}$ is the \emph{Cartan torsion tensor} \cite{Hehl1976_fundations,Gasperini-DeSabbata,Medina2018}. This last term represents the  geometrical counterpart of the spin inside the matter, which, along with the mass, fulfils a dynamical role in the EC framework. Hereafter, a hat symbol refers to quantities framed in GR. The affine connection $\Gamma^{\lambda}_{\mu \nu}$ can be also written as $\Gamma^\lambda_{\mu \nu}:=\hat{\Gamma}^{\lambda}_{\mu \nu}-K_{\mu \nu}^{\phantom{\mu \nu} \lambda}$, where $K_{\mu \nu}^{\phantom{\mu \nu} \lambda}$ is the \emph{contortion tensor}.  The EC field equations assume the GR-like form  
\begin{align} 
\hat{G}^{\alpha\beta}&=\frac{8\pi G}{c^4}\left(T^{\alpha\beta}+\frac{8\pi G}{c^4}\mathcal{S}^{\alpha \beta}\right),\label{eq:hat-G-equals-tilde-T}
\end{align}
where $T^{\alpha \beta}$ is the metric energy-momentum tensor, while $\mathcal{S}^{\alpha \beta}$, which we may dub \emph{\qm{torsional stress-energy tensor}}, depends on the \emph{spin angular momentum tensor} $\tau_{\gamma}^{\phantom{\gamma}\beta \alpha}$ (see Eq. (5c) in Ref. \cite{Paper2}).

The Weyssenhoff semiclassical model pertains to the description of a neutral spinning perfect fluid within   EC theory \cite{Obukhov1987,Boehmer2006}. First of all, the fluid is characterized by the spin angular momentum tensor 
\begin{align}
\tau_{\alpha\beta}{}^\gamma&=s_{\alpha\beta}u^\gamma,
\label{eq:spin-tensor-fluid}
\end{align}
where $s_{\alpha\beta}=s_{[\alpha\beta]}$ and $u^\alpha$ represent the spin density tensor  and the timelike four-velocity vector of the fluid, respectively.  Furthermore, it is subject to  the \emph{Frenkel condition} 
\begin{equation} \label{eq:Frenkel_condition}
 \tau_{\alpha\beta}{}^\beta= s_{\alpha\beta}\,u^\beta=0,
\end{equation}
which, in turn, leads to  the identity \cite{Paper2}
\begin{align} \label{eq:gauge_EC}
    S^{\alpha \mu}_{\phantom{\alpha \mu}\mu}=0.
\end{align}
Moreover, the metric and the torsional stress-energy tensors  are, respectively, \cite{Paper2}
\begin{align} 
T^{\alpha \beta} & = e \dfrac{u^\alpha u^\beta}{c^2}+ \mathcal{P}^{\alpha\beta} P 
\nonumber \\
& + 2 \left(\dfrac{u_\mu u^\gamma}{c^2}-\delta^\gamma_\mu\right)\hat{\nabla}_\gamma\left[s^{\mu(\alpha}u^{\beta)}\right] 
\nonumber  \\
& - \dfrac{16 \pi G}{c^4} \left(s^2 u^\alpha u^\beta + c^2 s^{\alpha}_{\phantom{\alpha}\lambda} s^{\beta \lambda}\right), \label{eq:T_alpha_beta_fluid} 
\\
\mathcal{S}^{\alpha \beta} &=2c^2 s^{\alpha}_{\phantom{\alpha}\lambda} s^{\beta \lambda} +s^2  u^\alpha u^\beta -\dfrac{1}{2}s^2 c^2 g^{\alpha \beta}, 
\label{eq:S-tensor-fluid}
\end{align}
where $e= \rho c^2 + \varepsilon$ is the fluid total energy density ($\rho$ and $\varepsilon$ being the rest-mass  and the internal energy densities, respectively), $\mathcal{P}^{\mu\nu}= \frac{u^\mu u^\nu}{c^2}+g^{\mu\nu}$  the projector operator on the hypersurface orthogonal to $u^\alpha$, $P$ the fluid pressure,  and $s^2 := s^{\alpha \beta}s_{\alpha \beta}$  the spin density scalar. 

The dynamics of the Weyssenhoff fluid is governed by a set of translational and  rotational equations \cite{Paper2,Obukhov1987}. The former is represented by the Euler equation 
\begin{align} \label{eq:translational_fluid_equation_2}
&  \mathcal{P}^\nu_\mu \partial_\nu P + \dfrac{1}{c^2} \left(P+ e \right) a_\mu - \dfrac{2}{c^2} \hat{\nabla}_\nu \left( u^\nu a^\rho s_{\rho \mu} \right) 
\nn \\
 &+\frac{16 \pi G}{c^4} a^\lambda s_{\lambda \rho} s_{\mu}^{\phantom{\mu}\rho}= - s_{\nu \rho} u^\sigma R_{\mu \sigma}^{\phantom{\mu \sigma}\nu \rho},
\end{align}
whereas the latter reads as 
\begin{align} \label{eq:rotational_fluid_equation}
    \hat{\nabla}_\lambda \left( s_{\mu \nu} u^\lambda \right)  & = \dfrac{a^\sigma}{c^2} \left(u_\mu   s_{ \sigma \nu}- u_\nu   s_{ \sigma \mu} \right),
\end{align}
where $a^\mu$ is the fluid  four-acceleration vector and $R_{\mu \sigma}^{\phantom{\mu \sigma}\nu \rho}$ the Riemann tensor.  Note that Eq. \eqref{eq:translational_fluid_equation_2} reduces to the GR Euler equation if the spin vanishes.  

\subsubsection{Post-Newtonian description}
\label{sec:PN_description}

The PN description of EC theory can be greatly simplified upon assuming that the torsion tensor has  vanishing trace (see Eq. \eqref{eq:gauge_EC}), as in this way it is possible to employ a harmonic gauge  having the same form as in GR \cite{Paper1,Paper2}. Therefore,  we can write  the 1PN-accurate metric tensor in terms of the Poisson-type potentials $U$ and $U_i$, and the superpotential $X$, which  are defined by, respectively, 
\begin{subequations}
\begin{align}
U\left(t,\boldsymbol{x} \right) &:= G  \int \dfrac{{\rm d}^3\boldsymbol{x}^\prime}{|\boldsymbol{x}-\boldsymbol{x}^\prime|}\, \sigma^\prime, \label{eq:U-potential-def}
\\
U_i\left(t,\boldsymbol{x} \right) &:= G  \int \dfrac{{\rm d}^3\boldsymbol{x}^\prime}{|\boldsymbol{x}-\boldsymbol{x}^\prime|}\, \sigma_i^\prime 
\label{eq:U-i-potential-def}
\\
 X \left(t,\boldsymbol{x} \right)&:= G\int \dd^3 \boldsymbol{x}^\prime \, |\boldsymbol{x}-\boldsymbol{x}^\prime| \sigma^{\prime} ,
\label{eq:superpotential-EC-definition}
\end{align}
\end{subequations}
where  the primed variables are evaluated at time $t$ and position $\boldsymbol{x}^\prime$ and 
\begin{subequations}
\label{eq:def_sigma-sigma_i}
\begin{align}
\sigma &:= \frac{T^{00}+T^{kk}}{c^2} + \frac{8 \pi G}{c^6}\left(\mathcal{S}^{00}+\mathcal{S}^{kk}\right), \\
\sigma_i &:= \frac{T^{0i}}{c} + \frac{8 \pi G}{c^5} \mathcal{S}^{0i}.
\end{align}
\end{subequations}
The metric energy-momentum tensor $T_{\mu\nu}$ admits the same PN structure as in GR  (see e.g. Eqs. (9.1.42)--(9.1.44) in Ref. \cite{Weinberg1972}).  Moreover, starting from the PN expansion of the spin angular momentum tensor $\tau_{\lambda}^{\phantom{\lambda}\mu \nu}$, it is possible to build the PN series of  the torsional stress-energy tensor $\mathcal{S}^{\mu \nu}$; further details can be found in Refs.  \cite{Paper1,Paper2}. 

Bearing in mind the above premises, it is possible to construct the PN expansions of the main objects underlying the Weyssenhoff model. First of all, if we write the fluid four-velocity  as $ u^\mu = \frac{u^0}{c} \left(c,\boldsymbol{v}\right)$ (with $\boldsymbol{v} := {\rm d}\boldsymbol{x}/{\rm d}t$ the coordinate velocity), then it follows from Eqs. \eqref{eq:T_alpha_beta_fluid} and \eqref{eq:S-tensor-fluid}  that the PN form of $\sigma$ and $\sigma_i$ reads as
\begin{subequations}
\label{eq:sigma-sigmai-sigmaii-Weyseenhoff}
\begin{align}
\sigma&=\rho^{\star \star} + \rho_{{\rm v}}  -\dfrac{4}{c^2}\partial_k\left(s_{kl}v^l\right) + {\rm O}\left( c^{-4}\right),
\label{eq:sigma-Weyseenhoff}
\\
\sigma_i&=\rho^\star v^i-\partial_k s_{ki} +{\rm O}\left( c^{-2}\right),
\label{eq:sigma-i-Weyseenhoff}
\end{align}
\end{subequations}
where we have defined
\begin{align}
\rho^{\star \star} & := \rho^\star \left[1+\dfrac{1}{c^2} \left(\dfrac{v^2}{2}+ \Pi -\dfrac{U}{2}\right)\right],   
\\
\rho_{\rm v}& := \dfrac{1}{c^2} \rho^\star \left(v^2 -\dfrac{U}{2}+ \dfrac{3P}{\rho^\star}\right),
\end{align}
with $\Pi := \varepsilon/\rho$ the  specific internal energy and   $\rho^\star := \frac{u^0}{c} \sqrt{-g} \rho = \rho + \OO\left(c^{-2}\right)$ the coordinate rest-mass density of the fluid, which, in turn,   satisfies the exact conservation equation 
\begin{align} \label{eq:continuity-eq-rho-star}
\dfrac{{\rm d}}{{\rm d}t} \rho^\star +\rho^\star \partial_k v^k=0,
\end{align}
where $\frac{{\rm d}}{{\rm d}t} f(t,\boldsymbol{x})= \partial_t f + v^k \partial_k f$. We note that in deriving Eq. \eqref{eq:sigma-sigmai-sigmaii-Weyseenhoff} we have exploited the Frenkel condition \eqref{eq:Frenkel_condition} and the fact that 
\begin{align}
s_{ij}={}^{(1)}s_{ij} + \OO\left(c^{-2}\right),   
\label{eq:PN-s-ij-and-s-ij-1}
\end{align}
${}^{(n)}s_{\mu \nu}$ denoting a factor going like $\frac{\bar{M} \bar{v}^n}{\bar{d}^2 c^{n-1}}$  ($\bar{M}$, $\bar{v}$, and $\bar{d}$ are the typical mass, internal velocity, and dimension of the source, respectively).

By virtue of Eqs. \eqref{eq:U-potential-def} and \eqref{eq:sigma-Weyseenhoff}, the instantaneous potential $U$ can be written as
\begin{align} \label{eq:U-potential-PN-form}
U = \hat{\mathscr{U}} + \dfrac{1}{c^2} \left(\hat{\psi} + \Sigma\right) + \OO\left(c^{-4}\right),
\end{align}
where we have adopted the following definitions\footnote{Although we have indicated the potential \eqref{eq:lowercase-psi-potential} with a hat symbol, we recall that the specific internal energy $\Pi$ receives contributions also from the spin tensor. Despite that, we will see that $\hat{\psi}$ assumes the same functional form as in GR.}:
\begin{subequations}
\label{eq:potentials-U-psi-Sigma}
\begin{align}
\hat{\mathscr{U}}\left(t,\boldsymbol{x}\right) &:=G  \int \dfrac{{\rm d}^3\boldsymbol{x}^\prime}{|\boldsymbol{x}-\boldsymbol{x}^\prime|}\rho^{\star \prime},
\label{eq:curly-U-potential-EC-theory}
\\
\hat{\psi} \left(t,\boldsymbol{x}\right) &:=G  \int \dfrac{{\rm d}^3\boldsymbol{x}^\prime}{|\boldsymbol{x}-\boldsymbol{x}^\prime|}\rho^{\star \prime} \left(\dfrac{3}{2} v^{\prime\, 2} -\hat{\mathscr{U}}^\prime + \Pi^\prime + \dfrac{3P^\prime}{\rho^{\star \prime}}\right), 
\label{eq:lowercase-psi-potential}
\\
\Sigma \left(t,\boldsymbol{x}\right) &:= 4G \int \dd^3 \boldsymbol{x}^\prime \dfrac{(x - x^\prime)_k}{|\boldsymbol{x}-\boldsymbol{x}^\prime|^3} s^\prime_{kl} v^{\prime\,l}.
\label{eq:Sigma-potential-EC}
\end{align}
\end{subequations}
Furthermore, as a consequence of   Eqs. \eqref{eq:U-i-potential-def} and \eqref{eq:sigma-i-Weyseenhoff}, we find for the potential $U_i$ that
\begin{align} \label{eq:U-i-potential-PN-form}
U_i= \hat{\mathscr{U}}_i +  \Sigma_i + \OO\left(c^{-2}\right),
\end{align}
where
\begin{subequations}
\label{eq:potentials-U-i-Sigma-i}
\begin{align}
\hat{\mathscr{U}}_i \left(t,\boldsymbol{x}\right) &:=G  \int \dfrac{{\rm d}^3\boldsymbol{x}^\prime}{|\boldsymbol{x}-\boldsymbol{x}^\prime|}\rho^{\star \prime} v^{\prime i},
\label{eq:curly-U-i-potential-EC-theory}
\\    
\Sigma_i \left(t,\boldsymbol{x}\right) &:= G \int \dd^3 \boldsymbol{x}^\prime \dfrac{(x - x^\prime)_k}{|\boldsymbol{x}-\boldsymbol{x}^\prime|^3} s^\prime_{ki}.
\label{eq:Sigma-i-potential-EC}
\end{align}
\end{subequations}
For the superpotential, we have from Eqs. \eqref{eq:superpotential-EC-definition} and \eqref{eq:sigma-Weyseenhoff}
\begin{align}
 X= \hat{\chi} + \OO\left(c^{-2}\right),
\end{align}
where 
\begin{align} \label{eq:potential-chi}
\hat{\chi} \left(t,\boldsymbol{x}\right):= G\int \dd^3 \boldsymbol{x}^\prime \,|\boldsymbol{x}-\boldsymbol{x}^\prime| \rho^{\star \prime}.
\end{align}
In order to to work out the potentials \eqref{eq:Sigma-potential-EC} and \eqref{eq:Sigma-i-potential-EC}, 
we have exploited  the divergence theorem jointly with the hypothesis according to which the spin density tensor $s_{\mu \nu}$ has compact support in the region occupied by the gravitational source. 

The PN dynamics of the fluid is obtained by expanding Eqs. \eqref{eq:translational_fluid_equation_2} and \eqref{eq:rotational_fluid_equation}.  At 1PN level,  the Euler equation \eqref{eq:translational_fluid_equation_2} yields,  after some calculations,  
\begin{align}
&\rho^\star \left(\dfrac{{\rm d}v^i}{{\rm d}t} - \partial_i \hat{\mathscr{U}} \right) + \partial_i P + \dfrac{1}{c^2} \Biggl[ v^i \partial_t P - \partial_i P \Biggl( \hat{\mathscr{U}} + \dfrac{v^2}{2} 
\nn \\    
&+ \dfrac{P + \varepsilon}{\rho^\star} \Biggr) \Biggr] + \dfrac{\rho^\star}{c^2} \Biggl[ \partial_i \hat{\mathscr{U}} \left(-v^2 + 4 \hat{\mathscr{U}}\right) + v^i \left( 4 v^k \partial_k \hat{\mathscr{U}}\right.
\nn \\
&\left.+3 \partial_t \hat{\mathscr{U}}\right) -4 \partial_t   \hat{\mathscr{U}}_i + 4v^j \left(\partial_i  \hat{\mathscr{U}}_j -\partial_j  \hat{\mathscr{U}}_i\right) - \partial_i \hat{\Psi} \Biggr]
\nn \\
& +\dfrac{\rho^\star}{c^2} \Biggl[-\partial_i \Sigma -4 \partial_t   \Sigma_i + 4v^j \left(\partial_i  \Sigma_j -\partial_j \Sigma_i\right) \Biggr]
\nn \\
&+ \dfrac{2}{c^2} \dfrac{ s_{ki}}{\rho^{\star}} \Biggl[ \partial_t \partial_k P + \partial_j \left(v^j \partial_k P \right)\Biggr]
\nn \\
& +\dfrac{2}{c^2} s_{jk}\Biggr{[}-v^k\partial_i\partial_j \hat{\mathscr{U}} +v^l\left(\delta_{i[k}\partial_{j]}\partial_l+\delta_{l[j}\partial_{k]}\partial_i\right) \hat{\mathscr{U}}
\nn \\
&+8 \pi G \partial_{[k} s_{i|j]} +2\partial_i \partial_{[j} \hat{\mathscr{U}}_{k]}+2\partial_i \partial_{[j} \Sigma_{k]}+\delta_{i[k}\partial_{j]}\partial_t \hat{\mathscr{U}}\Biggr{]}
\nn \\
&={\rm O}\left(c^{-4}\right),
\label{eq:1PN-Euler-equation-explicit-2}
\end{align}
where  we have exploited the Frenkel condition and we have  defined
\begin{align} \label{Eq:Capital-PSI-potential}
    \hat{\Psi}:= \hat{\psi} +\dfrac{1}{2}\partial^2_t \hat{\chi}.
\end{align}
Note that both the  terms involving the product between the spin tensor $s_{jk}$ and   the second order derivatives of the potentials, and  those depending on the factors $s_{jk} \partial_{[k} s_{i|j]}$ are due to the contribution of the Riemann tensor occurring on the right-hand side of Eq. \eqref{eq:translational_fluid_equation_2}.  For our purposes, we will need the leading-order expansion of the rotational equation, which, owing to Eq. \eqref{eq:rotational_fluid_equation}, is 
\begin{align}
\dfrac{{\rm d}}{{\rm d}t} s_{ij}+s_{ij}\partial_k v^k={\rm O}\left(c^{-2}\right).
\label{eq:continuity-equation-s-ij}
\end{align}
 
\subsection{$N$-body problem as the point-particle limit of the continuous description} 
\label{sec:PPL_Continuous}
In this section, we derive the equations  governing the 1PN-accurate dynamics of a system of $N$ gravitationally interacting bodies by performing the point-particle limit of Eq. \eqref{eq:1PN-Euler-equation-explicit-2}, which is outlined  in Sec. \ref{sec:PP_limit}. Then, after having worked out the explicit form of the potentials in Sec. \ref{sec:IN_EX_POTENTIALS},  we analyze the new EC spin-dependent terms occurring in Eq. \eqref{eq:1PN-Euler-equation-explicit-2} in Sec. \ref{sec:spin-terms-in-Euler-equation}. The analysis of the derivatives of the external potentials, contained in  Sec. \ref{sec:Derivative-ext-pot}, allows us to  finally derive  the desired 1PN  equations of motion  (see Sec. \ref{sec:equations_of_motion}). 

The point-particle procedure  is applied to a framework where the fluid distribution can be broken into a collection of $N$ separated components, usually referred to as bodies  \cite{Poisson-Will2014,Blanchet-Schafer1989}. The main advantage of this pattern consists in the fact that  Eq. \eqref{eq:1PN-Euler-equation-explicit-2}, which in general  comprises a set of partial and integro-differential equations, is transformed into  a set of ordinary differential equations, which thus can be more easily dealt with. 

The terms occurring in   Eq. \eqref{eq:1PN-Euler-equation-explicit-2} which are independent of the spin give rise to the well-known EIH equations (see chapter 9 of Ref. \cite{Poisson-Will2014} for details). These can be obtained in two equivalent ways : (1) by employing the point-particle limit and noting that the   angular momentum of each body, which stems from a macroscopic rotation,  vanishes in our framework; (2) by supposing that the fluid is made of  $N$ structureless point particles. In the latter case,   the mass density $\rho^\star$, being defined in terms of the Dirac delta function,  is assigned a distributional nature  and the ensuing divergent integrals are then regularized by means of either the Hadamard or the dimensional regularization prescription.   

In EC theory, the evaluation of the new spin-dependent part of Eq. \eqref{eq:1PN-Euler-equation-explicit-2} via the Dirac-delta formalism runs into some difficulties. Indeed,  the terms involving the products between the second-order spatial derivatives of the potentials and the spin density $s_{jk}$ (which assumes a distributional nature in this approach) yield  factors quadratic in the Dirac delta function, which are ill-defined in the Schwartz theory of distributions \cite{Lieb(2001)} (a formal method to handle the multiplication of distributions is provided by Colombeau theory \cite{Colombeau1984,Colombeau1985})\footnote{We note that the product of distributions arises in many research fields, such as  electrodynamics and particle physics \cite{Gsponer2008a,Gsponer2008b,Gsponer2008c}, and   gravitational shock-waves \cite{Dray1984,Battista_Riemann_boosted}.}. In this paper, we will circumvent these issues  by working out the spin-dependent terms occurring in the Euler equation \eqref{eq:1PN-Euler-equation-explicit-2} via the abovementioned point-particle limit. This method does not entail the presence of Dirac delta distributions and the related singularities.

\subsubsection{The point-particle limit}
\label{sec:PP_limit}

As pointed out before, the Weyssenhoff fluid modelling the gravitational source is supposed to be split into $N$ separated pieces. Therefore, we can express the coordinate fluid density and the spin density as, respectively,
\begin{subequations}
\begin{align}
\rho^\star &= \sum_A \rho^\star_A,
\\
s_{ik}&= \sum_A s^A_{ik},
\end{align}
\end{subequations}
where both $\rho^\star_A$ and $s^A_{ik}$ are nonvanishing only within the volume occupied by the body $A$. Hereafter, the bodies  and all their related quantities are indicated  with capital Latin indices $A,B,C=1, \dots, N$.  

It is convenient to define the following variables:
\begin{subequations}
\begin{align}
m_A &:= \int_A  \dd^3 \boldsymbol{x} \; \rho^\star,
\label{eq:material-mass-A}
\\
\varepsilon_{jki}s_A^i(t) &:=\int_A {\rm d}^3 \boldsymbol{x} \, s_{jk},
\label{eq:spin-vector-body-A}
\\
\boldsymbol{x}_A(t)&:=\dfrac{1}{m_A} \int_A \dd^3 \boldsymbol{x} \;  \rho^\star \boldsymbol{x}, 
\\
\boldsymbol{v}_A(t)&:=\dfrac{\dd \boldsymbol{x}_A}{\dd t}=\dfrac{1}{m_A} \int_A \dd^3   \boldsymbol{x} \;  \rho^\star \boldsymbol{v}, 
\\
\boldsymbol{a}_A(t)&:=\dfrac{\dd \boldsymbol{v}_A}{\dd t}=\dfrac{1}{m_A} \int_A \dd^3 \boldsymbol{x} \; \rho^\star \frac{\dd \boldsymbol{v}}{\dd t},
\end{align}
\end{subequations}
representing the material mass, the spin vector, the center of mass, the center of mass velocity, and the center of mass acceleration of $A$, respectively. Note that the domain of integration is independent of time and extends beyond the volume occupied by $A$. Owing  to the continuity equation \eqref{eq:continuity-eq-rho-star}, the material mass \eqref{eq:material-mass-A} is constant, whereas Eq.  \eqref{eq:continuity-equation-s-ij} implies that the spin vector \eqref{eq:spin-vector-body-A}  is conserved modulo $\OO\left(c^{-2}\right)$ corrections. Moreover, the following notations will be employed:
\begin{align}
\boldsymbol{d}_A &:=  \boldsymbol{x} - \boldsymbol{x}_A,  \qquad \quad \; \boldsymbol{n}_{A} := \dfrac{\boldsymbol{d}_{A}}{d_{A}},
\nn\\
\boldsymbol{r}_{AB} &:= \boldsymbol{x}_A - \boldsymbol{x}_B, \qquad  \boldsymbol{n}_{AB} := \dfrac{\boldsymbol{r}_{AB}}{r_{AB}}.
\end{align}

The (conserved) total mass-energy of the body $A$ is \cite{Paper2}
\begin{align}
M_A  &=\int_A {\rm d}^3 \boldsymbol{x} \, \rho^\star \left[1+\dfrac{1}{c^2} \left(\dfrac{w^2}{2} + \Pi -\dfrac{U_A}{2}\right)\right] 
\nonumber \\    
&+ {\rm O}(c^{-4}),
\label{eq:total-mass-bodyA}
\end{align}
$U_A$ being  the internal selfgravity of $A$ (further details will be given in Sec. \ref{sec:IN_EX_POTENTIALS} below),  and 
\begin{subequations}
\begin{align}
y^i &:= x^i -x^i_A\left(t\right), \label{eq:y-i-A-def}
\\
w^i &:= \dfrac{{\rm d}}{{\rm d}t} y^i = v^i - v^i_A\left(t\right),
\label{eq.w-i-A-def}
\end{align}
\end{subequations}
denoting  the position  relative to the center of mass $x^i_A$ and the velocity  relative to the  body velocity $v^i_A$ of a fluid element, respectively. 

Hereafter, we will exploit the following reasonable hypotheses regarding the bodies, which are supposed to be: (1) reflection symmetric about their center of mass; (2) in stationary equilibrium; (3) mutually well separated.
The stationary-equilibrium condition implies that any fluid element has vanishing velocity relative to the center of mass, namely in our calculations the terms involving $w^i$ can be ignored (see Eq. \eqref{eq.w-i-A-def}). Note that this hypothesis resembles the static equilibrium used in GR. The subsequent calculations, performed \qm{as in the GR static equilibrium case}, are not spoiled by the presence of the spin as long as the intrinsic rotation of each fluid element is stationary, i.e.,  the spin vector associated to each fluid element neither changes direction nor  varies in time \cite{Moller1962,Romano2019}. This relativistic issue presents already at the classical level, when we deal with \emph{(nonclosed) micropolar continuous systems}, which find  physical applications in ferromagnetic substances or liquid crystals \cite{Romano2014}.

If the bodies are well separated, then  $\ell_A/d_A \ll 1$, $\ell_A$ denoting the typical linear  dimension of  $A$. 
\begin{figure*}[ht!]
    \centering
    \includegraphics[scale=0.38]{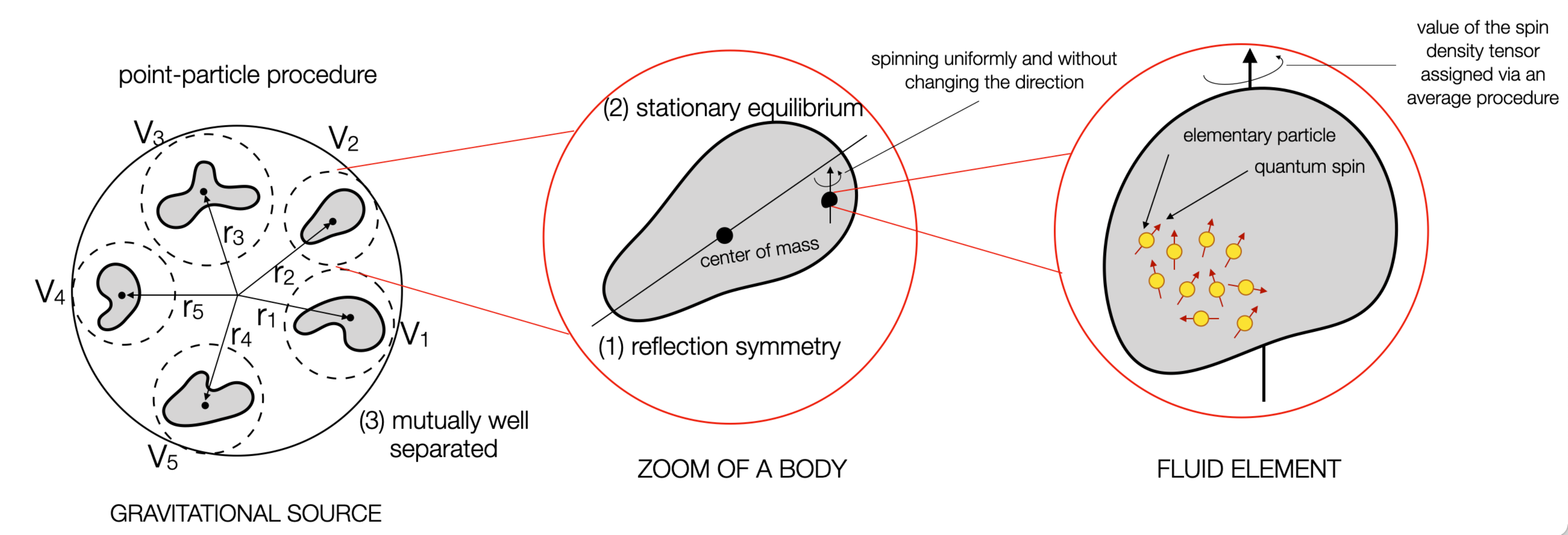}
    \caption{A pictorial sketch explaining the point-particle limit  and the hypotheses underlying  our  calculations, where the bodies are: (1) reflection symmetric about their center of mass; (2) in stationary equilibrium; (3) mutually well separated. In the Weyssenhoff fluid framework, the value of the spin density tensor of each fluid element, which contains a set of microscopic spin configurations, is assigned via an average procedure. Each fluid element is  spinning uniformly without changing direction. }
    \label{fig:Fig1}
\end{figure*}
For this reason, hereafter terms of fractional order $(\ell_A/d_A)^2$ or $(\ell_A/r_{AB})^2$ will be neglected. The hypotheses underlying our approach are visually summarized in Fig. \ref{fig:Fig1}. 

\subsubsection{The potentials}
\label{sec:IN_EX_POTENTIALS}

The potentials \eqref{eq:potentials-U-psi-Sigma}, \eqref{eq:potentials-U-i-Sigma-i}, and \eqref{eq:potential-chi} can be divided into internal and external pieces. The former represent the potentials produced by the body $A$,  while the latter refer to the potentials sourced by the remaining bodies of the system. Let $\mathscr{F}\left(t,\boldsymbol{x}\right) = \int \dd^3 \boldsymbol{x}^\prime \, f\left(t,\boldsymbol{x},\boldsymbol{x}^\prime \right) $ be a generic potential where the function $f$ has a compact support consisting of $N$ mutually disjoint connected regions. The internal and external contributions of $\mathscr{F}$ have the  form
\begin{align}
\mathscr{F}_A \left(t,\boldsymbol{x}\right) &= \int_A \dd^3 \boldsymbol{x}^\prime \, f\left(t,\boldsymbol{x},\boldsymbol{x}^\prime \right),
\nn \\
\mathscr{F}_{\neg A} \left(t,\boldsymbol{x}\right) &= \sum_{B \neq A} \int_B \dd^3 \boldsymbol{x}^\prime \, f\left(t,\boldsymbol{x},\boldsymbol{x}^\prime \right),
\end{align}
respectively,  so that $\mathscr{F}$ can be written  as
\begin{align}
    \mathscr{F}=  \mathscr{F}_A +  \mathscr{F}_{\neg A}.
\end{align}
For example,  the potential $\Sigma$ can be decomposed as (see Eq. \eqref{eq:Sigma-potential-EC})
\begin{subequations}
\begin{align}
\Sigma_A \left(t,\boldsymbol{x}\right) &= 4G \int_A \dd^3 \boldsymbol{x}^\prime \, s^\prime_{kl} \dfrac{(x - x^\prime)_k}{|\boldsymbol{x}-\boldsymbol{x}^\prime|^3}  v^{\prime\,l}  
\\
\Sigma_{\neg A}  \left(t,\boldsymbol{x}\right) &= \sum_{B \neq A} 4G \int_B \dd^3 \boldsymbol{x}^\prime \, s^\prime_{kl} \dfrac{(x - x^\prime)_k}{|\boldsymbol{x}-\boldsymbol{x}^\prime|^3} v^{\prime\,l}.  
\end{align}
\end{subequations}

In our hypotheses, the  potentials $\hat{\mathscr{U}}$, $\hat{\mathscr{U}}_i$, $\hat{\Psi}$ (cf. Eqs. \eqref{eq:curly-U-potential-EC-theory},  \eqref{eq:curly-U-i-potential-EC-theory}, and \eqref{Eq:Capital-PSI-potential}) assume the same form as in GR \cite{Poisson-Will2014}. In particular, we have
\begin{align}
\hat{\mathscr{U}}(t,\boldsymbol{x}) &= \sum_A \frac{G m_A}{d_A},
\\
\hat{\mathscr{U}}_i (t,\boldsymbol{x})&=\sum_A \frac{G m_A v_A^i}{d_A}.
\end{align}
Bearing in mind Eqs. \eqref{eq:Sigma-potential-EC}  and \eqref{eq:Sigma-i-potential-EC}, and adopting the same techniques as in GR (see e.g. chapter 9 of Ref. \cite{Poisson-Will2014}), for the spin-dependent potentials we find
\begin{align}
\Sigma (t,\boldsymbol{x}) &= 4G \sum_A \left(\boldsymbol{v}_A \times \boldsymbol{s}_A\right) \cdot \frac{\boldsymbol{n}_A}{d_A^2},
\label{eq:potential-Sigma-pp-limit}
\\
\Sigma_i (t,\boldsymbol{x}) &= G \sum_A \frac{\left(\boldsymbol{s}_A \times \boldsymbol{n}_A\right)^i}{d_A^2}.
\label{eq:potential-Sigma-i-pp-limit}
\end{align}

\subsubsection{Analysis of the spin-dependent terms}
\label{sec:spin-terms-in-Euler-equation}

In this section, we  work out the contributions involving the spin density $s_{ki}$ which occur in Eq. \eqref{eq:1PN-Euler-equation-explicit-2}.  In the following calculations, all functions inside integrals involving $y^i$ variables are supposed to depend on $t$ and $\boldsymbol{y}+\boldsymbol{x}_A(t)$. 

If we define the \emph{inner-structure-dependent quantity}
\begin{align}
\mathcal{H}^{ki}_A&:= 3G\int_A \dd^3 \boldsymbol{y} \, \dd^3 \boldsymbol{y}^\prime \, \rho^\star s^\prime_{kj} \frac{(y-y^\prime)^{\langle i}(y-y^\prime)^{j \rangle}}{\vert \boldsymbol{y}-\boldsymbol{y}^\prime \vert^5},
\label{eq:tensor-mathcal-H-A-ki}
\end{align}
then for the first group of spin-dependent terms appearing in Eq. \eqref{eq:1PN-Euler-equation-explicit-2},  we find
\begin{subequations}
\begin{align}
\int_A \dd^3 \boldsymbol{x} \, \rho^\star \partial_i \Sigma &= m_A \partial_i \Sigma_{\neg A} \left(t,\boldsymbol{x}_A \right) + 4 v^l_A \mathcal{H}^{li}_A,
\\
\int_A \dd^3 \boldsymbol{x} \, \rho^\star \partial_t\Sigma_i &=m_A \partial_t \Sigma_{i,\neg A}  \left(t,\boldsymbol{x}_A \right) -v^j_A  \mathcal{H}_A^{ij} 
\nn \\
&+ \OO\left(c^{-2}\right),
\label{eq:integral-2}
\\
\int_A \dd^3 \boldsymbol{x} \, \rho^\star v^j \partial_i\Sigma_j &=m_A v^j_A \partial_i \Sigma_{j,\neg A}  \left(t,\boldsymbol{x}_A \right) + v^j_A \mathcal{H}_A^{ji},
\end{align}
\end{subequations}
where we have exploited the continuity equation \eqref{eq:continuity-equation-s-ij} to derive Eq. \eqref{eq:integral-2}. Moreover, the spin-dependent quantities involving the second-order derivatives of the potentials give
\begin{subequations}
\begin{align}
\int_A \dd^3 \boldsymbol{x} \, s_{jk} v^k \partial_i \partial_j \hat{\mathscr{U}} &= \left(\boldsymbol{v}_A \times \boldsymbol{s}_A\right)^j \partial_i \partial_j \hat{\mathscr{U}}_{\neg A}\left(t,\boldsymbol{x}_A\right) 
\nn \\
& -v^k_A \mathcal{H}_A^{ki}, 
\\
\int_A \dd^3 \boldsymbol{x} \, s_{jk}  \partial_i \partial_j \hat{\mathscr{U}}_k &=\varepsilon_{jkl} s^l_A \partial_i \partial_j \hat{\mathscr{U}}_{k,\neg A}\left(t,\boldsymbol{x}_A\right) -v^k_A \mathcal{H}_A^{ki}, 
\\
\int_A \dd^3 \boldsymbol{x} \, s_{jk}  \partial_i \partial_j \Sigma_k &= \varepsilon_{jkl} s^l_A  \partial_i \partial_j \Sigma_{k,\neg A}\left(t,\boldsymbol{x}_A\right),
\\
\int_A \dd^3 \boldsymbol{x} \, s_{ji}  \partial_j \partial_t \hat{\mathscr{U}} &=\varepsilon_{jil} s^l_A  \partial_j \partial_t \hat{\mathscr{U}}_{\neg A} \left(t,\boldsymbol{x}_A\right) + v^k_A \mathcal{H}_A^{ik}.
\end{align}
\end{subequations}
Last, both the  integral
\begin{align}
&\int_A \dd^3\boldsymbol{x} \left(s_{jk} \partial_k s_{ij}-s_{jk} \partial_j s_{ik}\right) =2 \int_A \dd^3\boldsymbol{x}\,  s_{ij} \partial_k s_{kj},
\end{align}
and those involving the derivatives of the pressure  vanish owing to the reflection symmetry condition. 

\subsubsection{Derivatives of the external potentials}
\label{sec:Derivative-ext-pot}

In our hypotheses, the derivatives of $\hat{\mathscr{U}}_{\neg A}$ and  $\hat{\mathscr{U}}_{j,\neg A}$ assume the same form as in GR. In particular (see chapter 9 of Ref. \cite{Poisson-Will2014}), 
\begin{subequations}
\begin{align}
\partial_i \partial_j \hat{\mathscr{U}}_{\neg A}   \left(t, \boldsymbol{x}_A\right)&= \sum_{B \neq A} \frac{3Gm_B}{r_{AB}^3}n_{AB}^{\langle ij \rangle},
\\
\partial_i \partial_j \hat{\mathscr{U}}_{k,\neg A}   \left(t, \boldsymbol{x}_A\right)&= \sum_{B \neq A} \frac{3Gm_B}{r_{AB}^3}n_{AB}^{\langle ij \rangle}v^k_B,
\\
\partial_i \partial_t \hat{\mathscr{U}}_{\neg A}   \left(t, \boldsymbol{x}_A\right)&= \sum_{B \neq A} \frac{-3Gm_B}{r_{AB}^3}n_{AB}^{\langle ij \rangle}v_B^j.
\end{align}
\end{subequations}
For the new spin-dependent potentials,  we have
\begin{subequations}
\begin{align}
\partial_i \Sigma_{\neg A} \left(t, \boldsymbol{x}_A\right)&=\sum_{B \neq A} \frac{G}{r_{AB}^3}
\Bigl[  4\left(\boldsymbol{v}_B \times \boldsymbol{s}_B\right)^i 
\nn \\
&-12 \boldsymbol{n}_{AB} \cdot \left(\boldsymbol{v}_B \times \boldsymbol{s}_B\right)n_{AB}^i\Bigr],
\\
\partial_t \Sigma_{i,\neg A} \left(t, \boldsymbol{x}_A\right)&= \sum_{B \neq A}  \frac{G}{r_{AB}^3}\Bigl[\left(\boldsymbol{v}_B \times \boldsymbol{s}_B\right)^i
\nn \\
&- 3\left(\boldsymbol{n}_{AB} \times \boldsymbol{s}_B\right)^i \left(\boldsymbol{v}_B \cdot \boldsymbol{n}_{AB}\right)    \Bigr] 
\nn \\
&+\OO\left(c^{-2}\right),
\label{eq:EC-derivative-external-potential-2}
\\
\partial_i \Sigma_{j,\neg A} \left(t, \boldsymbol{x}_A\right)&= \sum_{B \neq A}  \frac{G}{r_{AB}^3}\Bigl[  3 n_{AB}^i \left(\boldsymbol{n}_{AB} \times \boldsymbol{s}_B\right)^j 
\nn \\
&+ \varepsilon_{ijl}s^l_B \Bigr],
\\
\partial_i\partial_j \Sigma_{k,\neg A} \left(t, \boldsymbol{x}_A\right)&= \sum_{B \neq A}\frac{G}{r_{AB}^4}\Biggl\{ 3  \Bigl[\delta_{ij}  \left(\boldsymbol{n}_{AB} \times \boldsymbol{s}_B\right)^k 
\nn \\
&+ \varepsilon_{kil} s_B^l n_{AB}^j + \varepsilon_{kjl} s_B^l n_{AB}^i  \Bigr] 
\nn  \\
&-15n_{AB}^i n_{AB}^j  \left(\boldsymbol{n}_{AB} \times \boldsymbol{s}_B\right)^k 
\Biggr\},
\end{align}
\end{subequations}
where  in Eq. \eqref{eq:EC-derivative-external-potential-2} we have exploited the continuity equation \eqref{eq:continuity-equation-s-ij}.

\subsubsection{Equations of motion}
\label{sec:equations_of_motion}
By means of the calculations of the previous sections, the coordinate acceleration $a_A^i$ of the body $A$ reads as 
\begin{align}
m_A a_A^i &= m_A a_{A,{\rm EIH}}^i   +  \frac{1}{c^2} \Biggl\{m_A \biggl[ \partial_i \Sigma_{\neg A} + 4 \partial_t \Sigma_{i,\neg A} -4 v^j_A
\nn \\
&\times  \left(\partial_i \Sigma_{j,\neg A}-\partial_j \Sigma_{i,\neg A}\right) \biggr] -2 \biggl[  2 \varepsilon_{jkl} s^l_A \partial_i \partial_j \hat{\mathscr{U}}_{k,\neg A}
\nn \\
&-2 \left(\boldsymbol{v}_A \times \boldsymbol{s}_A\right)^j \partial_i \partial_j \hat{\mathscr{U}}_{\neg A} + \varepsilon_{jik} s^k_A v^l_A \partial_j \partial_l \hat{\mathscr{U}}_{\neg A}  
\nn \\
&+2 \varepsilon_{jkl} s^l_A \partial_i \partial_j \Sigma_{k,\neg A} + \varepsilon_{jil} s^l_A \partial_j \partial_t \hat{\mathscr{U}}_{\neg A}  \biggr] \Biggr\} 
\nn \\
&+ \OO\left(c^{-4}\right),
\label{eq:EC-body-A-equation-of-motion-1}
\end{align}
where $a_{A,{\rm EIH}}^i$ is the EIH acceleration of the object $A$ and all the external potentials are evaluated at $\boldsymbol{x}=\boldsymbol{x}_A$. Bearing in mind the results of Sec. \ref{sec:Derivative-ext-pot}, the final form of the equations of motion for the body $A$ is
\begin{align} 
a_A^i &=a_{A,{\rm EIH}}^i + \frac{4}{c^2}\sum_{B \neq A} \frac{G}{r_{AB}^3} \Biggl\{ 2 \Bigl[ \left(\boldsymbol{v}_B-\boldsymbol{v}_A\right) \times \boldsymbol{s}_B\Bigr]^i  
\nn \\
&+3 n_{AB}^i \; \boldsymbol{s}_B \cdot  \left[\boldsymbol{n}_{AB} \times \left(\boldsymbol{v}_A-\boldsymbol{v}_B \right) \right]
\nn \\
&+3\left( \boldsymbol{n}_{AB} \times \boldsymbol{s}_B\right)^i \left(\boldsymbol{v}_A - \boldsymbol{v}_B\right) \cdot \boldsymbol{n}_{AB} \Biggr\}
\nn \\
&-\frac{6}{c^2} \sum_{B \neq A} \frac{G M_B}{M_Ar_{AB}^3} \Biggl\{  \Bigl[ \left(\boldsymbol{v}_A-\boldsymbol{v}_B\right) \times \boldsymbol{s}_A\Bigr]^i  
\nn \\
&-2 n_{AB}^i \; \boldsymbol{s}_A \cdot  \left[\boldsymbol{n}_{AB} \times \left(\boldsymbol{v}_A-\boldsymbol{v}_B \right) \right]
\nn \\
&+\left( \boldsymbol{n}_{AB} \times \boldsymbol{s}_A\right)^i \left(\boldsymbol{v}_B - \boldsymbol{v}_A\right) \cdot \boldsymbol{n}_{AB} \Biggr\}
\nn \\
&-\frac{12}{c^2} \sum_{B \neq A} \frac{G }{M_Ar_{AB}^4} \Biggl \{ s^i_A \left(\boldsymbol{n}_{AB}\cdot \boldsymbol{s}_B\right)+s^i_B \left(\boldsymbol{n}_{AB}\cdot \boldsymbol{s}_A\right)
\nn \\
&+ n_{AB}^i \Bigl[ \boldsymbol{s}_A \cdot \boldsymbol{s}_B - 5 \left(\boldsymbol{n}_{AB}\cdot \boldsymbol{s}_A\right)\left(\boldsymbol{n}_{AB}\cdot \boldsymbol{s}_B\right)\Bigr] \Biggr\}
\nn \\
&+ \OO\left(c^{-4}\right),
\label{eq:EC-body-A-equation-of-motion-2}
\end{align}
where  we have taken into account that  $M_A = m_A + \OO\left(c^{-2}\right)$ (see Eq. \eqref{eq:total-mass-bodyA}).  Equation \eqref{eq:EC-body-A-equation-of-motion-2}, jointly with the conservation law   $\dd \boldsymbol{s}_A / \dd t = {\rm O}\left(c^{-2}\right)$, completely determines the dynamics of the $N$-body system at 1PN level. 

From the above equations,  it is clear that, remarkably, the contributions of the tensor \eqref{eq:tensor-mathcal-H-A-ki} vanish identically. Furthermore, the external potentials do not couple with structure-dependent integrals (such as the mass multipole moments of the bodies) and their derivatives are written in terms of the bodies'  mass and  spin. In particular, Eq. \eqref{eq:EC-body-A-equation-of-motion-2} involves the total mass $M_A$ and not  its decomposition (see Eq. \eqref{eq:total-mass-bodyA}), and the spin of $A$ enters only via the definition \eqref{eq:spin-vector-body-A}. In other words, no corrections stemming  from the inner details of the bodies occur in the equations of motion at 1PN order, which imply that both the mass and the spin can be seen as labels characterizing the objects. This result can be interpreted as a hint for the validity of the \emph{effacing  principle} of the internal structure in EC theory. Apart from the hypotheses  (1)--(3) (see Fig. \ref{fig:Fig1}), which resemble the   GR pattern, this achievement has been  obtained  by means of the Frenkel condition. This  is a crucial requirement as it gives physical significance to the Weyssenhoff model and, as consequence, to EC theory as well. 

\section{Binary systems}
\label{sec:binary_system}

We apply the results of the previous section to the case of binary systems. The relative acceleration in the barycentric frame is evaluated in Sec. \ref{sec:relative_acceleration}. Then,  we estimate the new EC  contributions to the GR motion in  Sec. \ref{sec:estimate_EC_to_GR}. Last, we conclude the section with an  interesting analysis showing the conceptually close connections between GR and EC theories (see Sec. \ref{sec:similarities_GR_EC}).

\subsection{The relative acceleration}
\label{sec:relative_acceleration}

The relative dynamics of the two bodies can be readily described by defining the vectors\footnote{In this section, $\boldsymbol{v}$ is the relative velocity of the binary system and must not be confused with the fluid velocity field. }
\begin{align}
\boldsymbol{r} &:=\boldsymbol{x}_1-\boldsymbol{x}_2, \qquad \qquad \;\;\;  \boldsymbol{n}:= \boldsymbol{r}/r,
\nn \\
\boldsymbol{v} &:=\frac{\dd }{\dd t} \boldsymbol{r}=\boldsymbol{v}_1-\boldsymbol{v}_2, \qquad \boldsymbol{a} :=\frac{\dd }{\dd t} \boldsymbol{v}=\boldsymbol{a}_1-\boldsymbol{a}_2,
\label{eq:relative-vectors}
\end{align}
the spin variables
\begin{align}
\boldsymbol{s} := \boldsymbol{s}_1 + \boldsymbol{s}_2, \qquad \boldsymbol{\sigma} := \frac{M_2}{M_1}\boldsymbol{s}_1+\frac{M_1}{M_2}\boldsymbol{s}_2,
\end{align}
and the total mass $M$, the reduced mass $\mu$, and the symmetric mass ratio $\nu$ of the system  
\begin{align}
M & := M_1+M_2, \qquad \mu  := \frac{M_1M_2}{M}, \qquad \nu  := \frac{\mu}{M}.
\label{eq:total-Mass-et-al}
\end{align}

In a mass-centered coordinate system, the motion of the bodies is related to their relative motion by the following relations \cite{Paper2}: 
\begin{subequations}
\label{eq:position-vectors-r1-r2-with-spin}
\begin{align}
\boldsymbol{x}_1(t)&=\left[\frac{\mu}{M_1}+\frac{\mu (M_1-M_2)}{2M^2c^2}\left(v^2-\frac{GM}{r}\right)\right]\boldsymbol{r}(t)
\notag\\
&+\frac{2 \nu}{c^2}\left[\dfrac{\boldsymbol{s}_1(t)}{M_1} -\dfrac{\boldsymbol{s}_2(t)}{M_2}\right]\times \boldsymbol{v}(t)+{\rm O}\left(c^{-4}\right),
\\
\boldsymbol{x}_2(t)&=\left[-\frac{\mu}{M_2}+\frac{\mu (M_1-M_2)}{2M^2c^2}\left(v^2-\frac{GM}{r}\right)\right]\boldsymbol{r}(t)\notag\\
&+\frac{2 \nu}{c^2}\left[\dfrac{\boldsymbol{s}_1(t)}{M_1} -\dfrac{\boldsymbol{s}_2(t)}{M_2}\right]\times \boldsymbol{v}(t)+{\rm O}\left(c^{-4}\right).
\end{align}
\end{subequations}
Starting from Eq.  \eqref{eq:EC-body-A-equation-of-motion-2} with $N=2$ and employing the abovedefined quantities \eqref{eq:relative-vectors}-\eqref{eq:total-Mass-et-al}, the relative acceleration reads as 
\begin{align}
\boldsymbol{a} = \boldsymbol{a}_{\rm EIH} + \boldsymbol{a}_{\rm EC} + \OO \left(c^{-4}\right), 
\end{align}
where the GR contribution is 
\begin{align}
\boldsymbol{a}_{\rm EIH} &= -\frac{GM}{r^2} \boldsymbol{n} + \frac{GM}{c^2 r^2} \Biggl\{ \Bigl[ 2 (2 + \nu) \frac{GM}{r} + \frac{3}{2} \nu \left(\boldsymbol{n} \cdot \boldsymbol{v}\right)^2 
\nn \\
&- (1+3 \nu) v^2 \Bigr] \boldsymbol{n} + 2(2-\nu) \left(\boldsymbol{n} \cdot \boldsymbol{v}\right) \boldsymbol{v} \Biggr\},
\end{align}
whereas the EC correction is given by
\begin{align}
\boldsymbol{a}_{\rm EC} &=\frac{4G}{c^2r^3} \Biggl[ - \boldsymbol{v} \times \left(2 \boldsymbol{s} + \frac{3}{2}\boldsymbol{\sigma}\right) + 3 \boldsymbol{n} \left(\boldsymbol{n} \times \boldsymbol{v}\right) \cdot \left(\boldsymbol{s} + \boldsymbol{\sigma}\right)
\nn \\
&+ 3 \boldsymbol{n} \times \left(\boldsymbol{s}  +\frac{\boldsymbol{\sigma}}{2}\right)\left(\boldsymbol{n} \cdot \boldsymbol{v}\right) \Biggr] -\frac{12G}{c^2 r^4 \mu} \Biggl\{ \boldsymbol{s}_1 \left(\boldsymbol{n}\cdot \boldsymbol{s}_2\right)
\nn \\
&+ \boldsymbol{s}_2 \left(\boldsymbol{n}\cdot \boldsymbol{s}_1\right) + \boldsymbol{n} \Bigl[ \boldsymbol{s}_1 \cdot \boldsymbol{s}_2 -5 \left(\boldsymbol{n}\cdot \boldsymbol{s}_1\right) \left(\boldsymbol{n}\cdot \boldsymbol{s}_2\right) \Bigr] \Biggl\}.
\label{eq:EC-acceleration-binary}
\end{align}

The  last equation shows that the EC acceleration vector has the same  functional form as in GR. This result will be analyzed in Sec. \ref{sec:similarities_GR_EC}.  

\subsection{Numerical comparison with general relativity}
\label{sec:estimate_EC_to_GR}

We evaluate the EC contributions to the acceleration  by calculating the parameter $\epsilon:= \frac{|\boldsymbol{a}_{\rm EC}|}{|\boldsymbol{a}_{\rm EIH}| }$. We suppose that the bodies are black holes having masses $M_1=2M/3$, $M_2=M/3$,  relative radius  $\boldsymbol{r}=\left(100 GM/c^2,0,0\right)$, and relative  velocity $\boldsymbol{v}=\left(0,0.5\sqrt{GM/r},0\right)$. Following Ref. \cite{Paper2}, the spins can be modelled as  $\boldsymbol{s}_i=\frac{4\pi}{3} n \hbar  \left(\frac{2 G M_i}{c^2}\right)^3(0,0,1)$  ($i=1,2$), where $n= 10^{44}\, {\rm m}^{-3}$ is estimated as the inverse of the nucleon volume. In this  way, we find  $10^{-23} \lesssim \epsilon \lesssim 10^{-13} $  for $M\in[6,10^{11}]M_\odot$.

If the bodies have macroscopic angular momenta or \qm{classic spins} $\hat{\boldsymbol{s}}_1$ and $\hat{\boldsymbol{s}}_2$, then, after having defined
\begin{equation}
\hat{\boldsymbol{s}} := \hat{\boldsymbol{s}}_1 + \hat{\boldsymbol{s}}_2, \qquad \hat{\boldsymbol{\sigma}} := \frac{M_2}{M_1}\hat{\boldsymbol{s}}_1+\frac{M_1}{M_2}\hat{\boldsymbol{s}}_2,
\end{equation}
the GR relative acceleration can be written (in the center of mass frame) as 
\begin{align} \label{eq:GR-acceleration-with-SO-SS}
\boldsymbol{a}_{\rm GR}= \boldsymbol{a}_{\rm EIH}+\boldsymbol{a}_{\rm SO}+\boldsymbol{a}_{\rm SS}+ \OO \left(c^{-4}\right),    
\end{align}
where \cite{Poisson-Will2014} 
\begin{subequations}
\begin{align}
\boldsymbol{a}_{\rm SO}&= \frac{2G}{c^2r^3}\Biggl[-\boldsymbol{v}\times\left(2\hat{\boldsymbol{s}}+\frac{3}{2}\hat{\boldsymbol{\sigma}}\right)+3\boldsymbol{n}(\boldsymbol{n}\times\boldsymbol{v})\cdot(\hat{\boldsymbol{s}}+\hat{\boldsymbol{\sigma}})\notag\\
&+3\boldsymbol{n}\times\left(\hat{\boldsymbol{s}}+\frac{\hat{\boldsymbol{\sigma}}}{2}\right)(\boldsymbol{n}\cdot\boldsymbol{v})\Biggr],
\\
\boldsymbol{a}_{\rm SS} &=-\frac{3G}{c^2r^4\mu}\Biggl\{ \hat{\boldsymbol{s}}_1 \left(\boldsymbol{n}\cdot \hat{\boldsymbol{s}}_2\right)+ \hat{\boldsymbol{s}}_2 \left(\boldsymbol{n}\cdot \hat{\boldsymbol{s}}_1\right)
\nn \\
&+ \boldsymbol{n} \Bigl[ \hat{\boldsymbol{s}}_1 \cdot \hat{\boldsymbol{s}}_2 -5 \left(\boldsymbol{n}\cdot \hat{\boldsymbol{s}}_1\right) \left(\boldsymbol{n}\cdot \hat{\boldsymbol{s}}_2\right) \Bigr] \Biggr\}.
\end{align}
\end{subequations}
By employing the above equations, we can compute the EC contributions  via  the parameter  $\epsilon_{\rm spin}=|\boldsymbol{a}_{\rm EC}|/|\boldsymbol{a}_{\rm SO}+\boldsymbol{a}_{\rm SS}|$. We consider the same setup as before, while  for the \qm{classic spins} we write $\hat{\boldsymbol{s}}_i= \alpha \frac{GM_i^2}{c}(0,0,1)$ ($i=1,2$ and $\alpha \in (0,1)$). If   $\alpha = 1/2$, we obtain $10^{-20} \lesssim \epsilon_{\rm spin} \lesssim 10^{-10}$ with $M\in[6,10^{11}]M_\odot$.

\subsection{Links between general relativity and Einstein-Cartan theory}
\label{sec:similarities_GR_EC}

The analysis of the equations of motion performed in Sec. \ref{sec:relative_acceleration} reveals that, up to a redefinition of the spin variables, the 1PN-accurate EC and GR accelerations coincide (recall, however, the distinct nature featuring the quantum spin and the classical angular momentum).  Despite our starting point is represented by Eq. \eqref{eq:translational_fluid_equation_2}, which differs from the GR Euler equation, we find in fact that if 
\begin{align}\label{Eq:GR-EC-spin-related}
\hat{\boldsymbol{s}}\quad \leftrightarrow\quad   2 \boldsymbol{s}, 
\end{align}
then
\begin{align} \label{Eq:GR-EC-accel-related}
\boldsymbol{a}_{\rm SO} + \boldsymbol{a}_{\rm SS}\quad \leftrightarrow\quad \boldsymbol{a}_{\rm EC}.   
\end{align} 
Various explanations supporting Eq. \eqref{Eq:GR-EC-accel-related} can be provided.  First of all, the Frenkel condition \eqref{eq:Frenkel_condition} permits  to ignore, at 1PN level, all contributions stemming from the torsional stress-energy tensor \eqref{eq:S-tensor-fluid}. However, at higher PN orders, $\mathcal{S}^{\mu \nu}$  introduces additional corrections which can make  the EC acceleration  differ from  the GR one. Moreover, the terms appearing in Eq. \eqref{eq:1PN-Euler-equation-explicit-2}, which  involve the product between the spin and its first order derivatives and the derivatives of the pressure, vanish owing to the reflection symmetry (see Sec. \ref{sec:spin-terms-in-Euler-equation}). The result \eqref{Eq:GR-EC-accel-related} can  be also interpreted  by investigating the test-particle limit of the dynamical equations, where one body is nearly at rest while its companion has a small mass with a finite spin-to-mass ratio. In fact, within this approximation, the 1PN-accurate GR acceleration  agrees with the 1PN dynamics, as  described by the Mathisson-Papapetrou equations, of a test particle endowed with \qm{classic spin} in the background gravitational field of a Kerr black hole  \cite{Tanaka1996,Tagoshi2000,Faye2006a}. Although the motion of a spinning test particle in EC theory is described by a set of Mathisson-Papapetrou-like equations generalizing the aforementioned  equations valid in GR (see Eq. (8) in Ref. \cite{Hehl1971}), the test-mass limit of the EC and GR accelerations will lead to the same effects by virtue of Eq. \eqref{Eq:GR-EC-accel-related}. However, this is consistent with the following two facts: (1) we have checked that, within our hypotheses and at 1PN level, the EC Mathisson-Papapetrou-like equations reduce to the corresponding  GR equations; (2) if we employ the potentials \eqref{eq:potential-Sigma-pp-limit} and \eqref{eq:potential-Sigma-i-pp-limit} along with Eq. \eqref{Eq:GR-EC-spin-related}, the  metric, when evaluated for a single body having vanishing $\boldsymbol{x}_A$ and $\boldsymbol{v}_A$,   reproduces the 1PN Kerr metric in harmonic coordinates. This last  result goes in the direction of the findings of Ref. \cite{Arkuszewski1974}, where it has been proved that in the weak-field limit the metric tensor of a static body made of Weyssenhoff dust coincides with the linearized Kerr metric. 

\section{Discussion and conclusions} 
\label{sec:end}

In this paper we have investigated the $N$-body problem in EC theory at 1PN level by exploiting the Weyssenhoff fluid to model the spin effects inside matter. To achieve this objective, our methodology expounds on  the point-particle limit of the Weyssenhoff fluid's continuous description to finally derive the related equations of motion \eqref{eq:EC-body-A-equation-of-motion-2}, see Sec. \ref{sec:PPL_Continuous}. This procedure relies on three fundamental assumptions on each body, which are (see Fig. \ref{fig:Fig1}): (1) reflection symmetric about their center of mass; (2) in stationary equilibrium; (3) mutually well separated. During our calculations, we have proved the no-dependence of the equations of motion on structure-dependent terms. This is an essential clue for the validity of the effacing principle at 1PN order in EC theory, which states that the internal (gravitational) details of each extended body in the system do not influence its own dynamics as soon as hypothesis (3) holds. This permits also to avoid tidal effects among the objects, which surely spoil hypotheses (1) and (2) as well.

The Frenkel condition \eqref{eq:Frenkel_condition} provides a physical meaning to the Weyssenhoff fluid model, and leads to a drastic simplification of the ensuing calculations. More in general, this situation implies reflexively that assumption \eqref{eq:gauge_EC} in EC framework is vital to make the theory coherent. As one can observe, Eq. \eqref{eq:Frenkel_condition}
leads to a wealth of beneficial consequences not only in terms of purely mathematical and numerical computations (see Ref. \cite{Paper2}, for details), but also under conceptual perspectives. In fact, it can be exploited as a sort of criterion to select, among all possible EC models, those endowed with physical connotations. It would be interesting to investigate this particular class of EC theories and check whether, besides the effacing principle, the equivalence principle (in its various formulations) holds (see Ref. \cite{Dicasola2015} for a comprehensive review on the different formulations and meanings of the equivalence principle). However, something in this direction has already been proved by Von der Hyde \cite{Vonderhyde1975}. This topic fulfils a paramount task in building up solid extensions of GR, being also in agreement with its foundation principles.                

In Sec. \ref{sec:binary_system}, we have applied our findings to binary systems. We have numerically compared the EC spin contributions to the GR bulk dynamics via $\epsilon$, and then to the GR macroscopic angular momentum via $\epsilon_{\rm spin}$, obtaining thus $10^{-23} \lesssim \epsilon \lesssim 10^{-13}$ and  $10^{-20} \lesssim \epsilon_{\rm spin} \lesssim 10^{-10}$ for all black hole mass ranges $M\in[6,10^{11}]M_\odot$. The effect remains physically very small as soon as the bodies keep widely separated (or, in the gravitational-wave terminology, in the inspiral stage). Furthermore, we have discovered that at 1PN order the GR and EC treatments are conceptually equivalent up to a constant factor relating the quantum and \qm{classic} spins (cf. Eqs. \eqref{Eq:GR-EC-spin-related} and \eqref{Eq:GR-EC-accel-related}). Nevertheless, we strongly expect that such equivalence should break down at higher PN orders, because EC theory sprouts up on new terms (e.g., $\mathcal{S}_{\alpha\beta}$), stemming \emph{de facto} from its geometrical description (see conclusions of Ref. \cite{Paper2}, for similar discussions). 

Another important cross-checking theoretical result relies on having verified that at 1PN level the GR Mathisson-Papapetrou equation is surprisingly recovered also in EC theory, as the two approaches move their steps from essentially dissimilar hypotheses. Moreover,  we obtain, as in GR, the 1PN approximation of the Kerr metric. However, the latter result possesses two distinct physical interpretations in GR and EC frameworks, albeit they mathematically reproduce the same metric (up to a normalization factor). 

The outcomes of our paper can be compared with those obtained in the literature in the broad framework of general relativistic theories with torsion. In fact, the PN scheme has allowed the  authors of Refs. \cite{Schweizer1979,Smalley1980} to discover that GR and  teleparallel theories of gravitation (where curvature vanishes) agree at 1PN level, but differ at higher orders. The same conclusion holds also for the 1PN generation of the gravitational radiation, as discussed in Ref. \cite{Schweizer1980}. In particular, it is shown that the dipole catastrophe, which afflicts many alternative metric theories of gravity, is absent. The PN formalism has been applied also to EC theory by Castagnino and collaborators \cite{Castagnino1985,Castagnino1987}, who have employed the ideal spinning  fluid model to derive  the 1PN dynamical equations of a matter source and a test particle moving in the vacuum region outside the source distribution. In this approach, the study of the 1PN acceleration of the test particle can, in  principle, lead to the possibility of distinguishing GR and EC theories. This pattern differs from the one adopted in this paper, where we have employed the point-particle limit to describe a system of bodies subject to their mutual gravitational attraction and, in addition, our starting point is represented by the 1PN Euler equation \eqref{eq:1PN-Euler-equation-explicit-2}. We also mention the paper of Gladchenko \& Zhytnikov  \cite{Gladchenko1994}, who have considered the 1PN approximation of the quadratic Poincar\'e gauge theory  of gravitation in its most general form, where torsion quanta are allowed. Here, differently from our study, their existence must be constrained via classical gravity effects, like light deflection and time delay tests. Last, as more recent applications we point out some works on PN and parametrized PN (PPN) expansions performed in a general class of teleparallel gravity theories \cite{Ualikhanova2019,Emtsova2020,Gonzalez2022}. It emerges that the two PPN parameters $\beta$ and $\gamma$ allow to highlight the differences with GR, whereas in the limit of $f(T)$ theories ($T$ being the torsion scalar) indistinguishability with GR is again restored.

Our findings have shown that in EC theory all spin-related quantities come naturally out of the theory. Therefore, it might be interesting to calculate the  macroscopic angular momentum in GR by resorting to the EC pattern (similarly to the analysis of Refs. \cite{Ray1982a,Ray1982b}). The great advantage of this approach dwells in the possibility to carry consistently out the calculations, without unbinding the physical nuances. 

The further step after this study will be its Lagrangian formulation together with the analysis of the related first integrals. Although at 1PN order EC and GR accelerations coincide, important deviations are likely to emerge in the PN analysis of the rotational equation \eqref{eq:rotational_fluid_equation}. These topics will deserve consideration in a separate paper.

\section*{Acknowledgements}
The authors are grateful to Gruppo Nazionale di Fisica Matematica of Istituto Nazionale di Alta Matematica for partial support. E.B. acknowledges the support of the Austrian Science Fund (FWF) grant P32086. V.D.F. thanks Prof. Antonio Romano for the stimulating discussions on the internal angular momentum in continuous media. V.D.F. acknowledges the support of INFN {\it sezione di Napoli}, {\it iniziative specifiche} TEONGRAV.

\end{document}